\documentclass[10pt,twocolumn,letterpaper]{article}

\usepackage[pagenumbers]{cvpr} % To force page numbers, e.g. for an arXiv version

\usepackage{stfloats}
\usepackage{graphicx}
\usepackage{amsmath}
\usepackage{amssymb}
\usepackage{booktabs}
\usepackage{multirow}
\usepackage{makecell}
\usepackage{threeparttable}

\usepackage[pagebackref,breaklinks,colorlinks]{hyperref}

\usepackage[capitalize]{cleveref}
\crefname{section}{Sec.}{Secs.}
\Crefname{section}{Section}{Sections}
\Crefname{table}{Table}{Tables}
\crefname{table}{Tab.}{Tabs.}

\def\cvprPaperID{5776} % *** Enter the CVPR Paper ID here
\def\confName{CVPR}
\def\confYear{2022}

\begin{document}

\title{KFWC: A Knowledge-Driven Deep Learning Model for Fine-grained Classification of Wet-AMD}

\author{
    Haihong E\textsuperscript{1,3}, Jiawen He\textsuperscript{1,3}, Tianyi Hu\textsuperscript{1,3}, Lifei Wang\textsuperscript{2}, Lifei Yuan\textsuperscript{2}, Ruru Zhang\textsuperscript{1,3}, Meina Song\textsuperscript{1,3}\\
    \textsuperscript{1} Beijing University of Posts and Telecommunications \quad \textsuperscript{2} Hebei Eye Hospital\\
    \textsuperscript{3} Education Department Information Network Engineering Research Center \\ (Beijing University of Posts and Telecommunications)\\
    {\tt\small \{ehaihong, euphy, hutianyi, zrr, mnsong\}@bupt.edu.cn}
}

    
\maketitle

%%%%%%%%% ABSTRACT
\begin{abstract}
   Automated diagnosis using deep neural networks can help ophthalmologists detect the blinding eye disease wet Age-related Macular Degeneration (AMD). Wet-AMD has two similar subtypes, Neovascular AMD and Polypoidal Choroidal Vessels (PCV). However, due to the difficulty in data collection and the similarity between images, most studies have only achieved the coarse-grained classification of wet-AMD rather than a finer-grained one of wet-AMD subtypes. To solve this issue, in this paper we propose a Knowledge-driven Fine-grained Wet-AMD Classification Model (KFWC), to classify fine-grained diseases with insufficient data. With the introduction of a priori knowledge of 10 lesion signs of input images into the KFWC, we aim to accelerate the KFWC by means of multi-label classification pre-training, to locate the decisive image features in the fine-grained disease classification task and therefore achieve better classification. Simultaneously, the KFWC can also provide good interpretability and effectively alleviate the pressure of data collection and annotation in the field of fine-grained disease classification for wet-AMD. The experiments demonstrate the effectiveness of the KFWC which reaches 99.71\% in AU-ROC scores, and its considerable improvements over the data-driven w/o Knowledge and ophthalmologists, with the rates of 6.69\% over the strongest baseline and 4.14\% over ophthalmologists.
\end{abstract}

%%%%%%%%% BODY TEXT

\section{Introduction}
\begin{figure}[t]
  \centering
  \includegraphics[width=0.95\linewidth]{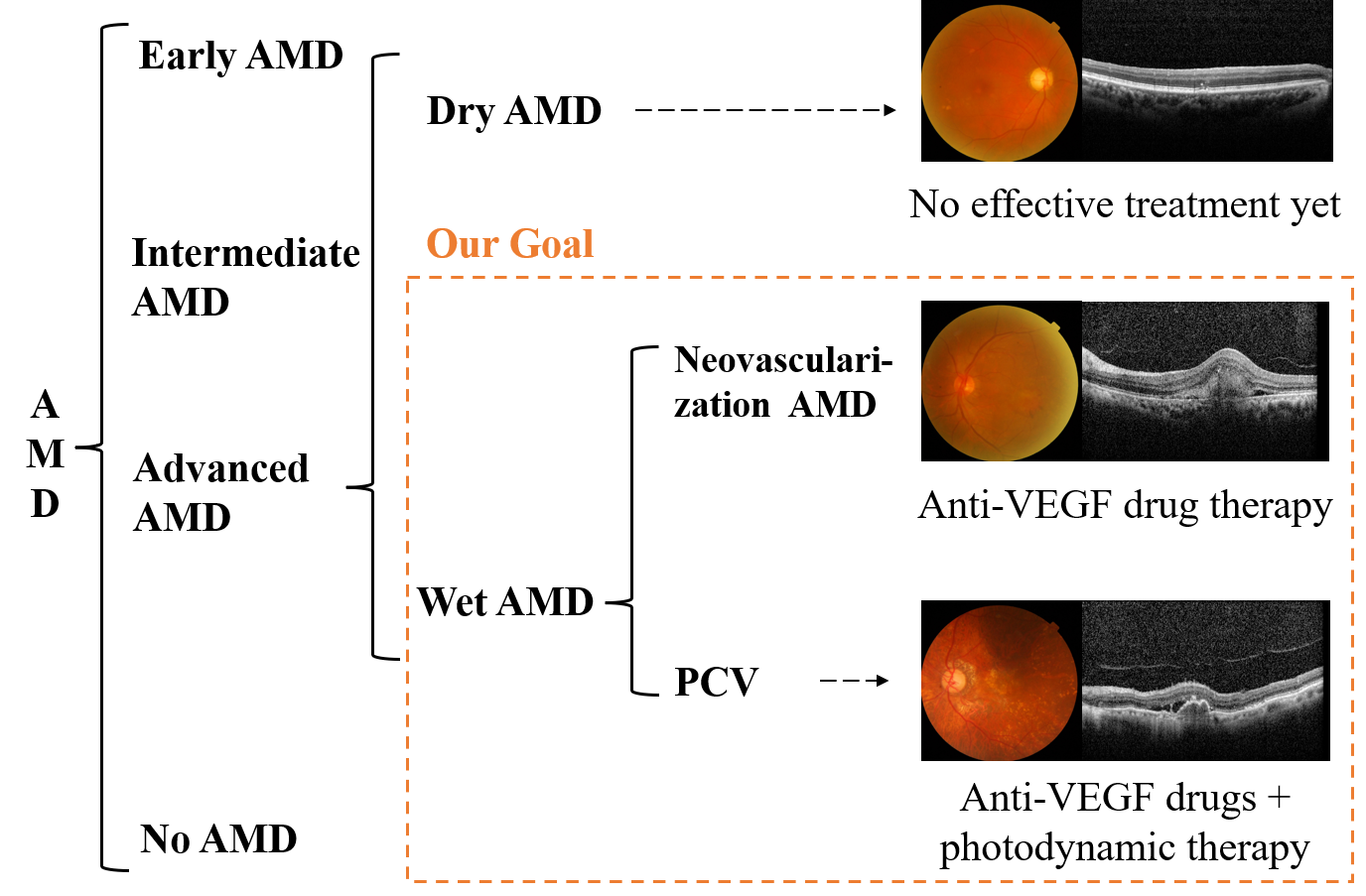}

   \caption{AMD Staging Standards}
   \label{fig:1}
\end{figure}

Age-related Macular Degeneration (AMD) is a common and serious eye disease. AMD causes approximately 9\% of blindness worldwide and is the leading cause of vision loss in developed countries\cite{quartilho2016leading}. As shown in Figure \ref{fig:1}, according to the AMD staging criteria from Age-Related Eye Disease Study (AREDS) \cite{davis2005age}, AMD can be divided into early AMD, intermediate AMD, and advanced AMD. Advanced AMD can be divided into dry-AMD and wet-AMD. Wet-AMD can be further differentiated into neovascular AMD and Polypoid Choroidal Vessels (PCV) \cite{wong2016age}. Unlike neovascular AMD, PCV needs to be treated with photodynamic therapy  besides anti-VEGF drugs \cite{cheung2018polypoidal}, which has significantly differences in clinical treatment. Therefore, it is necessary to accurately distinguish PCV from neovascular AMD in clinical practice.

Despite the importance of finer-grained classification between neovascular AMD and PCV, very few studies\cite{xu2021automated} have achieved that.  Most researches on AMD\cite{ismik20192019}\cite{hwang2019artificial}\cite{wang2019two}\cite{wang2020explainable} concentrate only on coarse-grained classification of AMD or wet-AMD. This is because fine-grained disease classification has two challenges as follows:

\begin{enumerate}
    \item
    % {\textbf{Data Scarcity}: Data-driven method is the key for many typical Convolutional Neural Network (CNN) researches (AlexNet\cite{krizhevsky2012imagenet}, Inception V3\cite{szegedy2016rethinking}, EfficientNet\cite{tan2019efficientnet} and so on), since they all rely on ImageNet\cite{deng2009imagenet}, a massive image dataset with annotation. However, this method is not applicable to fine-grained disease classification. On one hand, medical dataset is more difficult to collect and annotate; on the other hand, fine-grained task naturally leads to a small data base, which cause data hard to obtain.}
    
    {\textbf{Data Scarcity}: There is a significant data scarcity in the task of finer-grained classification between neovascular AMD and PCV, bacause fine-grained task naturally has a smaller data base compared to other coarse-grained tasks, which cause data hard to obtain. Therefore, typical data-driven method for many CNN research like AlexNet\cite{krizhevsky2012imagenet}, Inception V3\cite{szegedy2016rethinking}, EfficientNet\cite{tan2019efficientnet} and so on, is not applicable to fine-grained disease classification, since they all rely on ImageNet\cite{deng2009imagenet}, a massive image dataset with annotation.}

    \item 
    {\textbf{Image similarity}: CNN is widely considered as a black box, and the model performance is highly dependent on the processing ability of image features. In the analysis of medical images, image features include the size, shape and perspective of lesion regions\cite{gong2021deformable}. In the fine-grained disease classification task, different diseases are distinguished only at several key lesions, while most of the remaining image features are not helpful, which puts higher demands on the feature processing and feature localization capabilities of CNNs.}
\end{enumerate}

\begin{figure*}
  \centering
  \includegraphics[width=1\textwidth]{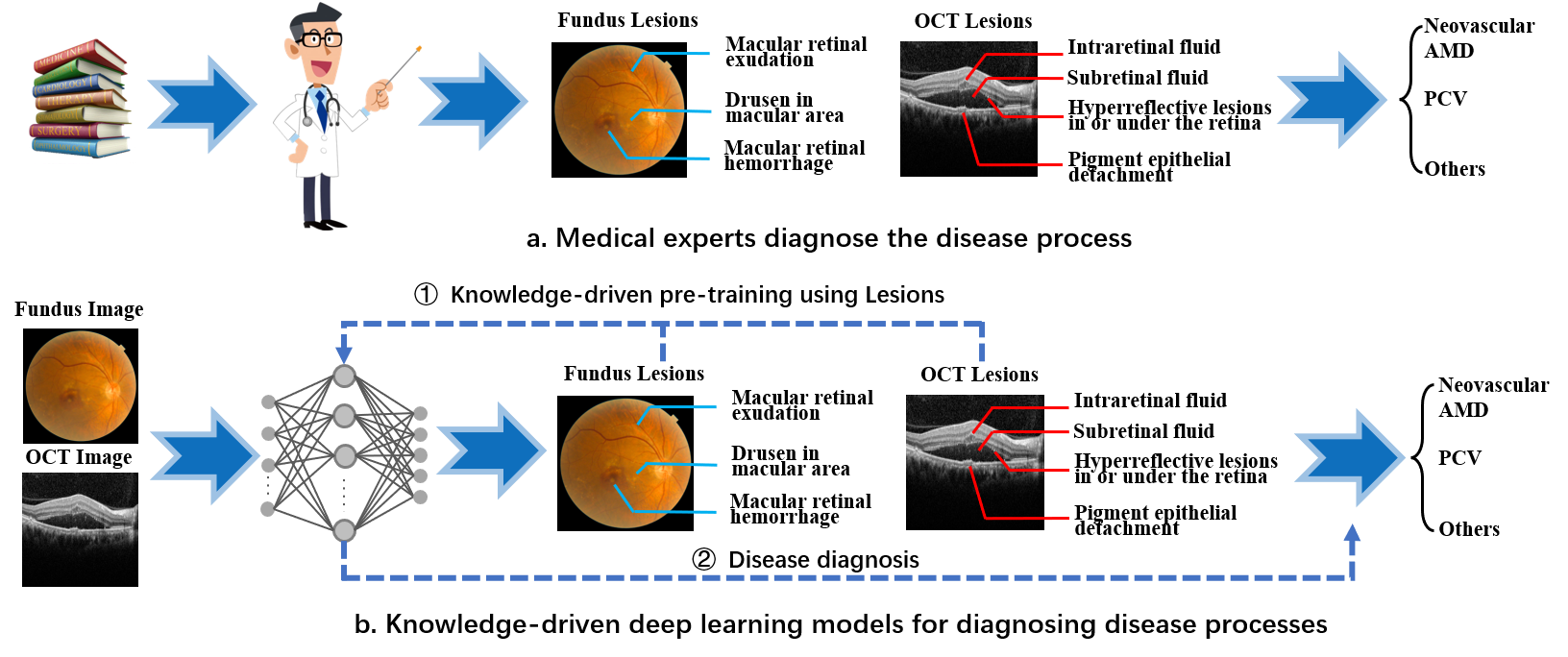}
  \caption {a. Human doctors gain knowledge from medical books, then use knowledge to analyze the lesion signs in fundus and OCT images, and finally diagnose the disease based on lesion signs. \\
  b. KFWC gains knowledge through pre-training of prior knowledge, then it uses knowledge of lesion signs to enhance the image feature processing and feature positioning capabilities, and finally achieves good results in disease classification tasks.}
  \label{fig:2}
\end{figure*}

To effectively address the two challenges in fine-grained disease classification mentioned above, in this paper we construct a fine-grained classification dataset of wet-AMD and proposes a Knowledge-driven Fine-grained Wet-AMD Classification Model (KFWC).
About the dataset, the dataset of wet-AMD consists of color fundus and OCT bi-modal images, and the images were annotated with 10 lesion signs as well as wet-AMD disease diagnosis by experts from a top hospital in China.
About the model, KFWC draws on the process of disease diagnosis by human doctors, as shown in Figure \ref{fig:2}. Doctors use what they have learned from a wealth of medical books to determine lesion signs in medical images and use them to make disease diagnoses. These images containing lesion signs can be regarded as a kind of knowledge, derived from the wisdom of medical experts, which can be used directly in the diagnosis of diseases. Similarly, we firstly   pre-train KFWC using multi-labeled lesion sign dataset as knowledge, and subsequently fine-tune it using disease diagnosis dataset. By artificially introducing a priori knowledge of lesion signs annotated by ophthalmologists, the ability of image feature processing and feature localization can be improved, the convergence of the model speeds up , and thus the dependence on the amount of data is reduced.

Our contributions are highlighted as follows:
\begin{itemize}
    \item
    {\textbf{Dataset}: We construct a fine-grained classification dataset of wet-AMD.}
    
    \item
    {\textbf{Model}: We propose the KFWC to enhance the accuracy of the model in fine-grained disease classification with insufficient data. We innovatively introduced a two-stage method. In the first stage, we introduce prior knowledge of 10 lesion signs through pre-training; in the second stage, the model realizes the classification task with both human knowledge and model independent learning.}
    
    \item
    {\textbf{Interpretability}: KFWC in the first stage can recognize 10 lesion signs, which provides interpretability for the classification conclusion of the model and also provides more auxiliary information for clinicians.}
    
    \item
    {\textbf{Experiments}: We conducted experiments with six common backbones, with the best model achieving an AUROC of 99.71\%, which has an improvement of 6.69\% over baseline and 4.14\% over ophthalmologists. Besides, heat map experiment shows KFWC can focus on the lesion sign more precisely.}
    
\end{itemize}

\section{Related Work}

\subsection{Data-driven Auxiliary Diagnosis Works}

Lee CS \etal~\cite{lee2017deep} proposed a deep neural network and used 101,002 OCT images to train it, which achieved the classification of normal and AMD. The AUC and accuracy reached 92.78\% and 87.63\%, respectively. Prahs P \etal~\cite{prahs2018oct} used 171,024 OCT images to train a CNN, and divided them into two categories based on whether anti-VEGF drugs needed to be injected. The classification accuracy of the model reached 95.5\%. Kermany D S \etal~\cite{kermany2018identifying} proposed a deep-learning framework and used 108,312 OCT images for training. This framework realized the classification of choroidal neovascularization, diabetic macular edema, drusen, and normal, and the AUC of which reached 99.96\%. Hwang D K \etal~\cite{hwang2019artificial} used 35,900 OCT images for CNNs training, and realized the classification of normal, dry-AMD, active wet-AMD, and inactive wet-AMD. The accuracy of the optimal CNN reached 92.67\%.

The above works have provided a large amount of data for the deep learning models, so that they could master the relevant knowledge. However, data-driven method has put great pressure on data acquisition and annotation, which is not applicable in the field of classification of neovascular AMD and PCV.

\subsection{Knowledge-driven Auxiliary Diagnosis Works}

Shukla K K \etal~\cite{shukla2017classification} used the image analysis software Fiji~\cite{schindelin2012fiji} to segment the pathological images and obtained 16 features such as tissue structure, cell distribution, and cell shape. After that, they used this knowledge to classify non-cancer cells and cancer cells and achieved an accuracy of 85.7\%. De Fauw J \etal~\cite{de2018clinically} proposed a knowledge-driven deep learning architecture for assisted diagnosis of fundus diseases. In the first stage, they segmented the OCT image, and the tissue segmentation map containing 15 kinds of labels was output. In the second stage, the segmentation map was used to realize the classification of diseases diagnosis and referral suggestions. This method obtains the prior knowledge of the tissue and lesion area through segmentation, and is superior to human experts in various indicators. Son HM \etal~\cite{son2021ai} used U-Net to segment the erythema area in the skin image and then cropped it according to the segmentation result. Then the cropped image was used for classification of 18 types of skin diseases. Experiments showed that after adding segmentation prior knowledge, the classification model obtained a higher accuracy rate.

In the field of auxiliary diagnosis, there are few researches use knowledge-driven, and most of them obtain prior knowledge by segmenting the lesion areas. Some studies have segmented a large number of tissues or lesion types, such as De Fauw J \etal~\cite{de2018clinically} This method can obtain a wealth of prior knowledge, but the difficulty and cost of data annotation will also greatly increase. Some studies only performed a single type of segmentation, such as Son HM \etal~\cite{son2021ai} This method can effectively relieve the pressure of data annotation, but the acquired prior knowledge will be very limited.

\section{Dataset}

 The images used in this paper are the real clinical images of a Grade III Level A ophthalmology hospital in China. The dataset contains 1,096 color fundus images and 4,927 OCT images, with high image clarity. All images are desensitized and do not contain any personal information of the patients. Four ophthalmologists with rich clinical experience annotated the dataset. They all have master degree in ophthalmology, including 2 chief physicians and 2 deputy chief physicians.
 
 The annotation is divided into two stages. In the first stage, the ophthalmologists separately annotate the lesion signs on the color fundus images and the OCT images, and each image gets 5 kinds of annotation results. In the second stage, the ophthalmologists synthesize the lesion signs on the two images and annotate the type of disease for the eye, namely neovascular AMD, PCV or others. Since the annotation of lesion signs and disease types have a medical logical order, we believe that lesion signs can be used as a kind of priori knowledge for disease classification.
 
\begin{figure*}
  \centering
  \begin{subfigure}{0.72\linewidth}
    \includegraphics[width=1\textwidth]{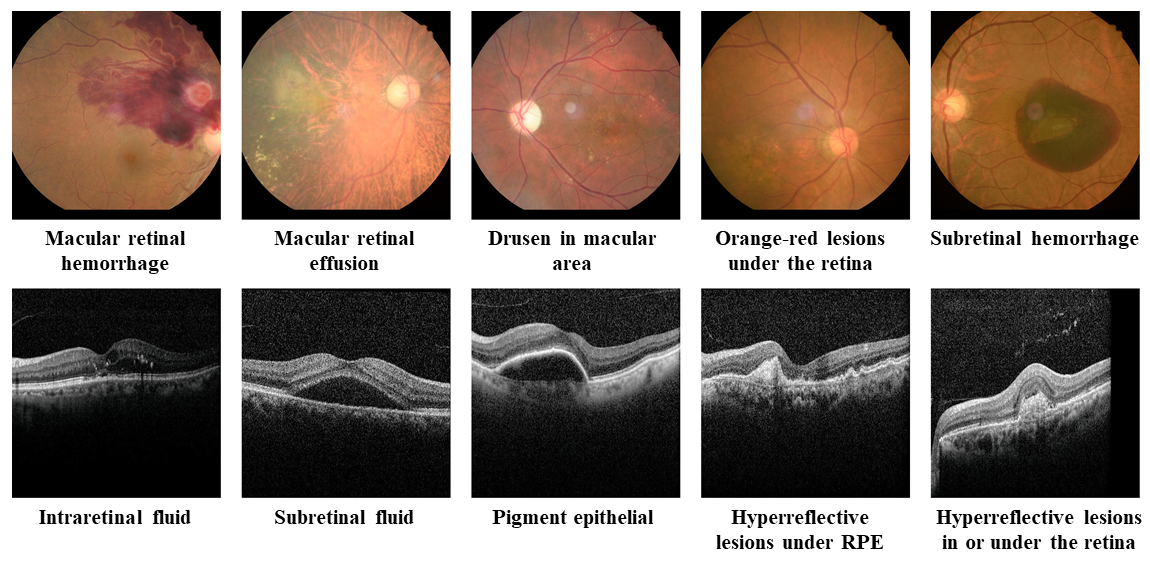}
    \caption{An example of lesion signs.}
    \label{fig:short-a}
  \end{subfigure}
  \begin{subfigure}{0.25\linewidth}
    \includegraphics[width=1\textwidth]{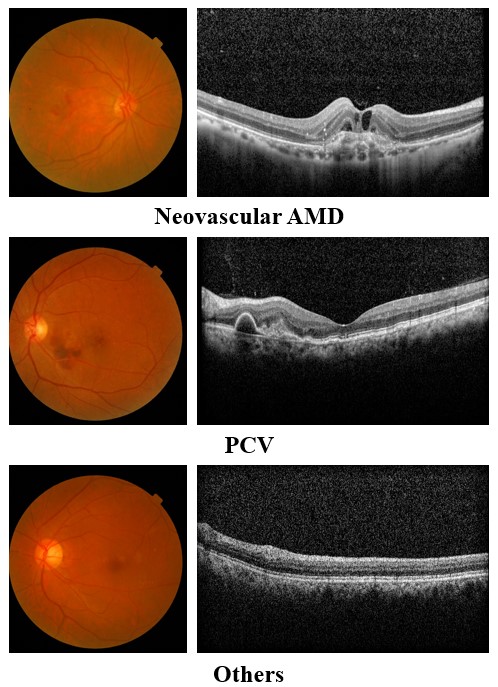}
    \caption{An example of bi-modal data.}
    \label{fig:short-b}
  \end{subfigure}
  \caption{Example of dataset.}
  \label{fig:3}
\end{figure*}

 We matched the color fundus images and OCT images of the same eye into one group, and a total of 5,261 groups of bi-modal data were obtained. An example of the data is shown in Figure \ref{fig:3}.
 
 The dataset division is shown in Table \ref{tab:1} and Table \ref{tab:2}. In the data preprocessing stage, the size of each image is modified to 224×224×3, and image enhancement operations such as random rotation, random horizontal flip, and random contrast enhancement are performed on the training dataset.

 \begin{table}[htbp]\footnotesize
  \centering
  \begin{tabular}{@{}lcccc@{}}
    \toprule
    \textbf{Type} & \textbf{Train} & \textbf{Valid} & \textbf{Test} & \textbf{Total} \\
    \midrule
    \textbf{Fundus image} & 882 & 107 & 107 & 1096  \\
    \textbf{OCT image} & 4072 & 364 & 491 & 4927 \\
    \bottomrule
  \end{tabular}
  \caption{Distribution of multi-label classification dataset of lesion signs.}
  \label{tab:1}
\end{table}

 \begin{table}[htbp]\footnotesize
  \centering
  \begin{tabular}{@{}lcccc@{}}
    \toprule
    \textbf{Type} & \textbf{Train} & \textbf{Valid} & \textbf{Test} & \textbf{Total} \\
    \midrule
    \textbf{Neovascular AMD} & 603 & 16 & 15 & 634 \\
    \textbf{PCV} & 1299 & 223 & 230 & 1752 \\
    \textbf{Others} & 2582 & 183 & 110 & 2875 \\
    \midrule
    \textbf{Total} & 4484 & 422 & 355 & 5261 \\
    \bottomrule
  \end{tabular}
  \caption{Distribution of bi-modal dataset.}
  \label{tab:2}
\end{table}
 
\section{Method}

\begin{figure*}
  \centering
  \includegraphics[width=1\textwidth]{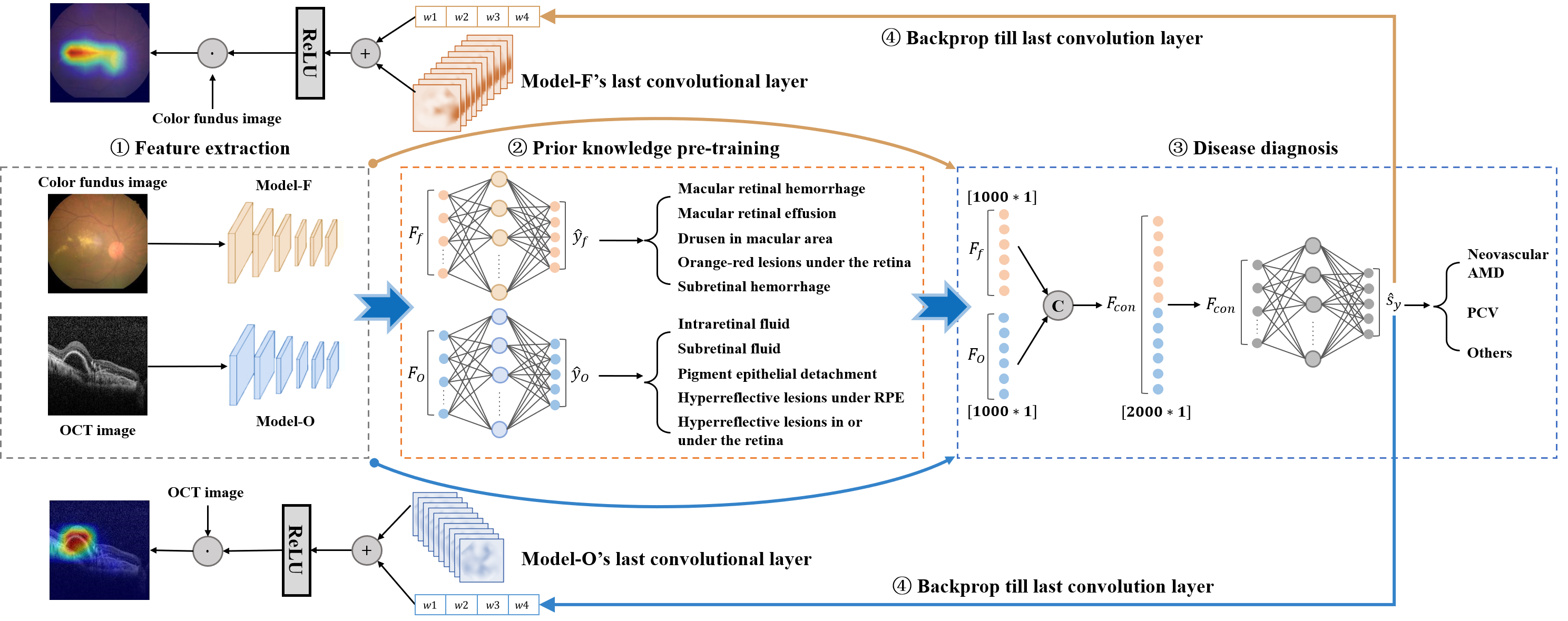}
  \caption{KFWC overview: First, the fundus image and OCT image were used for multi-label classification of lesion signs on two pipelines. After that, we reuse the feature extraction module and concat the two feature vectors to classify neovascular AMD, PCV, and others. Finally, we use Grad-CAM \cite{selvaraju2017grad} to generate visual explanations from KFWC.}
  \label{fig:4}
\end{figure*}

As mentioned earlier, in order to effectively introduce the prior knowledge of ophthalmologists, we propose a knowledge-driven two-stage deep learning model KFWC. KFWC receives color fundus images and OCT images commonly used in ophthalmological diagnosis, recognizes 10 common signs of wet-AMD on the images, and applies relevant knowledge to wet-AMD classification tasks. The interpretability and accuracy of disease diagnosis are effectively improved. The architecture of KFWC is shown in Figure \ref{fig:4}.

First, we define three datasets $D_f=\{x_f |y_f\}$, $D_O=\{x_O |y_O\}$ and $D=\{x_f,x_O |y\}$, where $x_f$ and $x_O$ represent color fundus image and OCT image respectively, $x_f,x_O \in R^{224\times 224 \times 3}$. $D_f$ and $D_O$ are lesion signs dataset for the color fundus and OCT images, respectively. $y_f$ is the annotation result of 5 common lesion signs of wet-AMD on fundus images, and $y_f\subset$ \{macular retinal hemorrhage, macular retinal exudation, drusen in macular area, orange-red lesions under the retina, subretinal hemorrhage\}. $y_O$ is the annotation result of 5 common lesion signs of wet-AMD on OCT images. $y_O\subset$ \{intraretinal fluid, subretinal fluid, pigment epithelial detachment, hyperreflective lesions under retinal pigment epithelium (RPE), hyperreflective lesions in or under the retina\}. The dataset $D$ is a bi-modal wet-AMD diagnostic dataset. Each group of $x_f$ and $x_O$ are color fundus image and OCT image obtained from the same eye, and $y$ is the diagnostic label of them, $y\in$ \{neovascular AMD, PCV, other\}. Our KFWC receives the input $\{x_f,x_O\}$, and outputs the diagnosis result $\hat{y}$ and the lesion sign recognition results $\hat{y}_f$ and $\hat{y}_O$:

\begin{equation}
  \hat{y},\hat{y}_f,\hat{y}_O \leftarrow KFWC(\{x_f, x_O\})
  \label{eq:1}
\end{equation}

\subsection{Feature Extraction}

The feature extraction module of KFWC consists of two symmetrical branches. Among them, the backbone used to process color fundus images is denoted as KFWC-F, and which used to process OCT images is denoted as KFWC-O. They can be any common CNNs, but for the consistency of subsequent operations, we will uniformly modify the output feature vectors $F_f$ and $F_O$ to 1000 dimensions, namely $F_f,F_O \in R^{1000}$.

In this paper, we use ResNet\cite{he2016deep}, Inception-v3\cite{szegedy2016rethinking}, SCNet\cite{liu2020improving} and ResNeSt\cite{zhang2020resnest} as candidate backbones to extract features.

\subsection{Prior Knowledge Pre-training}

In the first stage, we conduct pre-training on the prior knowledge of wet-AMD diagnosis, that is, multi-label classification training for 10 lesion signs. The pre-training process enables KFWC to quickly locate the features that need to pay attention to. The dependence on the amount of data can be effectively reduced, and the learning efficiency and accuracy can be improved.

The color fundus image and OCT image are two independent pipelines, as shown in Figure \ref{fig:4}. The color fundus image pipeline uses the dataset $D_f$ for training. $x_f$ is sent to KFWC-F and processed into a 1000-dimensional feature vector $F_f$, and then the score of the final output $\hat{y}_f$ is obtained through the fully connected layer. Similarly, we get $F_O$ and  $\hat{y}_O$ in the OCT image pipeline:

\begin{equation}
  \begin{cases}s_{\hat{y}_f} = W_f F_f \\ s_{\hat{y}_O} = W_O F_O \end{cases}
  \label{eq:2}
\end{equation}

where $W_f,W_O \in R^{1000 \times 5}$, $s_{\hat{y}_f},s_{\hat{y}_O} \in R^5$. By selecting the categories with scores greater than 0.5 in $s_{\hat{y}_f}$ and $s_{\hat{y}_O}$ , we get $\hat{y}_f$ and $\hat{y}_O$ in Equation \ref{eq:1}, that is, the signs of lesions on the color fundus images and OCT images.

In the pre-training process of prior knowledge, KFWC-F and KFWC-O will fine-tune the weights according to the results of multi-label classification of lesion signs. In this way, KFWC is forced to learn the corresponding relationship between image features and prior knowledge, then it can pay attention to image features that play a decisive role in the diagnosis of wet-AMD. At the same time, KFWC can output the recognition results of 10 kinds of lesion signs on bi-modal images, providing clinicians with abundant and reliable auxiliary information. The difficulty of multi-label classification annotation is far less than that of segmentation annotation. Therefore, compared with the existing knowledge-driven methods, our method reduces the cost of dataset annotation effectively.

\subsection{Wet-AMD Diagnosis}

As shown in Figure \ref{fig:4}, after the prior knowledge pre-training task is completed, we comprehensively consider bi-modal images and perform the diagnosis of two subtypes of wet-AMD, namely neovascular AMD and PCV.

At this stage, we use the dataset $D$ for training. We reuse KFWC-F and KFWC-O, and concat the feature vectors $F_f$ and $F_O$ to get the vector $F_{con} \in R^{2000}$. Finally, the score of the final output $\hat{y}$ is obtained through the fully connected layer:

\begin{equation}
  s_{\hat{y}}=W_{con} F_{con}
  \label{eq:3}
\end{equation}

where $W_{con}\in R^{2000\times3}$, $s_{\hat y}\in R^3$. The diagnosis of neovascular AMD and PCV is achieved by selecting the category with the highest score in $s_{\hat y}$.

At this stage, we reused the weights of the feature extraction model pre-trained. Instead of learning the image features of each disease from the beginning, KFWC can effectively use prior knowledge and focus on the signs of lesions during disease diagnosis. At the same time, the recognition results of lesion signs can also provide better interpretability for disease diagnosis. KFWC can cover the entire clinical diagnosis process of wet-AMD, effectively alleviating the pressure of clinicians in diagnosis and treatment. The effectiveness of KFWC will be verified experimentally in Chapter 5.

\subsection{Heat Map Visualization}

In order to visually explain the area of interest of KFWC during disease diagnosis, we utilize the Grad-weighted Class Activation Mapping (Grad-CAM) module \cite{selvaraju2017grad}. Grad-CAM can visualize CNNs of any structure without modifying the network structure or retraining. We mark the last convolutional layers of KFWC-F and KFWC-O as $A_f$ and $A_O$. In order to obtain the discriminant graph Grad-CAM corresponding to the disease category $c$, we backpropagate the corresponding scores $\hat{s_y}^c$ of the class $c$ output by the disease diagnosis module to $A_f$ and $A_O$. We perform global average pooling of the reflowed gradients in the width and height dimensions to obtain the importance weights of feature maps $w_f^c$ and $w_O^c$:

\begin{equation}
  \begin{cases} w_{fk}^c=\frac{1}{Z}\sum_i^h\sum_j^w \frac{ \partial \hat{ s_y}^c}{\partial A_{fij}^{k}} \\\\ w_{Ok}^c=\frac{1}{Z}\sum_i^h\sum_j^w \frac{ \partial \hat{ s_y}^c}{\partial A_{Oij}^{k}} \end{cases}
  \label{eq:4}
\end{equation}

where $h$ and $w$ are the height and width of the feature map of the last layer, $Z=h*w$. This weight captures the degree of influence of the channel $k$ of the feature map on the target category $c$. We add weighted activation values of feature maps, and then use ReLU to obtain a rough heat map:

\begin{equation}
  \begin{cases} L_{Grad-CAM_f}^c=ReLU(\sum_k w_{fk}^c A_f^k ) \\\\ L_{Grad-CAM_O}^c=ReLU(\sum_k w_{Ok}^c A_O^k ) \end{cases}
  \label{eq:5}
\end{equation}

\section{Experiments}
\subsection{Experimental Design}
(1) To select the most suitable backbone for wet-AMD diagnosis, we first designed wet-AMD lesion sign multi-label classification experiments on different backbone networks as a pre-training process. Based on these pre-trained models, we designed wet-AMD diagnostic experiments so as to select the best model and identify it as the network selected by KFWC. (2) To validate the performance of the KFWC, we replicate past studies with the same granularity and introduce the accuracy of ophthalmologists for comparison. (3) To validate the effectiveness of knowledge-driven, we designed ablation experiments to verify the improvements. (4) To explain how the knowledge works, we visualize heat maps for models with or without knowledge-driven so as to show how knowledge is transferred. These experimental results show that the prior Knowledge effectively improves the accuracy of KFWC and can identify the location of lesion signs more correctly and precisely.

\subsection{Implementation Details}
% ImageNet\cite{deng2009imagenet} provides a massive image dataset, and models trained with ImageNet can be used as starter models for other recognition tasks. We initialize the weight parameters of two backbones using the ImageNet pre-trained model from the PyTorch library and normalize our input dataset using the data distribution in ImageNet. Both backbones used for color fundus and OCT image feature extraction will remain the same type in all experiments.

While training the model, we use the SGD optimizer and take a dynamic learning rate approach for optimization. Learning rate is initially set to 0.001 and decreases every 20 epochs. The weight decay is set to 0.001, the momentum is set to 0.9, and the minimum batch size is 8. In the first stage, the model is trained for 500 epochs in order to fully acquire prior knowledge, using the BCE loss function. In the second stage, the sign extraction part of models from first stage is retained, after replacing the classifier, the model is trained for 100 epochs, using the cross-entropy loss function. All models are implemented based on the PyTorch framework and are performed on an NVIDIA GeForce GTX-1080Ti GPU with 12GB of RAM.

\subsection{Baseline Methods} \label{Baseline}
We compare the performance of KFWC with the state-of-the-art research and ophthalmologists.

1) Xu \etal~\cite{xu2021automated} proposed a bi-modal deep CNN framework that directly extracted features from color fundus images and OCT images by data-driven method, and used ResNet50 as the baseline. It is a representative of the data-driven approach at the same task. Since we can't get their dataset, we reproduced their work on the dataset constructed in this paper.

2) We calculate the diagnostic performance of ophthalmologists through 2 repeated rounds of disordered annotation. The first round of annotation result is used as the prediction of wet-AMD diagnosis and the second round of annotation result is used as the true wet-AMD diagnosis result. The diagnostic performance of ophthalmologists is obtained by calculating the predicted and true values. The result shows in Table \ref{tab:6}, line 'Ophthalmologists(average)' .

\subsection{Backbone Selection}

We use six feature extraction models for color fundus images and OCT images, respectively, and the experimental results of lesion signs multi-label classification experiments are shown in Table \ref{tab:3} and \ref{tab:4}. From the experimental results, we can see that the AUROC of color fundus images can reach above 0.9369, and the AUROC of OCT images can reach above 0.9223, therefore, the classification results are reliable. These experiments have validated the effectiveness of KFWC in lesion signs recognition and confirmed the reliability of knowledge. In addition, the recognition results can also demonstrate the diagnostic basis of the model to clinicians and patients, effectively enhance the interpretability of the model.

In order to select the best wet-AMD diagnostic model, we retained the feature extraction module weights obtained in the first stage of experiments and conducted wet-AMD diagnostic experiments. The results of wet-AMD diagnostic experiments under different backbones are listed in Table \ref{tab:5}. It can be seen that among the six backbones, the best performer is KFWC using ResNeSt50 as the feature extraction model, which reached 0.9971 in the AUROC index.

\begin{table}[htbp]\footnotesize 
  \setlength\tabcolsep{3pt}
  \begin{center}
        \begin{tabular}{lccccc}
            \toprule
            \textbf{Model} & \textbf{AUROC} & \textbf{Precision} & \textbf{Recall} & \textbf{F1} & \textbf{Acc}\\
            \midrule
            \textbf{KFWC-F(ResNet18)} & 0.9567 & 0.9058 & 0.8019 & 0.8474 & 0.7757 \\
            \textbf{KFWC-F(ResNet34)} & 0.9578 & 0.8995 & 0.8113 & 0.8499 & 0.8037 \\
            \textbf{KFWC-F(ResNet50)} & 0.9369 & 0.9064 & 0.7925 & 0.8425 & 0.7570 \\
            \textbf{KFWC-F(Inception-v3)} & 0.9378 & 0.8210 & 0.7453 & 0.7793 & 0.7383 \\
            \textbf{KFWC-F(SCNet50)} & 0.9639 & 0.8968 & 0.8491 & 0.8701 & 0.8224 \\
            \textbf{KFWC-F(ResNeSt50)} & 0.9445 & 0.8751 & 0.8208 & 0.8446 & 0.7757 \\
            \bottomrule
        \end{tabular}
        \caption{Multi-label classification experiments for AMD lesion signs in color fundus images}
        \label{tab:3}
    \end{center}
\end{table}  

\begin{table}[htbp]\footnotesize 
    \setlength\tabcolsep{3pt}
    \begin{center}
    \begin{tabular}{lccccc}
        \toprule
        \textbf{Model} & \textbf{AUROC} & \textbf{Precision} & \textbf{Recall} & \textbf{F1} & \textbf{Acc}\\
        \midrule
        \textbf{KFWC-O(ResNet18)} & 0.9269 & 0.7235 & 0.7443 & 0.7303 & 0.7060 \\
		\textbf{KFWC-O(ResNet34)} & 0.9261 & 0.7110 & 0.7748 & 0.7373 & 0.7033 \\
		\textbf{KFWC-O(ResNet50)} & 0.9223 & 0.7158 & 0.7634 & 0.7366 & 0.7088 \\
		\textbf{KFWC-O(Inception-v3)} & 0.9293 & 0.7327 & 0.7710 & 0.7476 & 0.7170 \\
		\textbf{KFWC-O(SCNet50)} & 0.9225 & 0.7182 & 0.7824 & 0.7467 & 0.7335 \\
		\textbf{KFWC-O(ResNeSt50)} & 0.9250 & 0.7519 & 0.7748 & 0.7624 & 0.7500 \\
		\bottomrule
    \end{tabular}
    \caption{Multi-label classification experiments for AMD lesion signs in OCT images}
    \label{tab:4}
  \end{center}
  
\end{table}

\begin{table}[htbp]\footnotesize 
  \setlength\tabcolsep{3pt}
  \begin{center}
    
    \begin{tabular}{lccccc}
        \toprule
        \textbf{Model} & \textbf{AUROC} & \textbf{Precision} & \textbf{Recall} & \textbf{F1} & \textbf{Kappa}\\
        \midrule
        \textbf{KFWC(ResNet18)} & 0.9888 & 0.9053 & 0.8409 & 0.8527 & 0.6605 \\
		\textbf{KFWC(ResNet34)} & 0.9775 & 0.8977 & 0.8494 & 0.8551 & 0.6967 \\
		\textbf{KFWC(ResNet50)} & 0.9797 & 0.9012 & 0.8608 & 0.8643 & 0.7405 \\
		\textbf{KFWC(Inception-v3)} & 0.9495 & 0.9023 & 0.8835 & 0.8907 & 0.6170 \\
		\textbf{KFWC(SCNet50)} & 0.9608 & 0.8556 & 0.8381 & 0.8406 & 0.5060 \\
		\textbf{KFWC(ResNeSt50)} & \textbf{0.9971} & \textbf{0.9373} & \textbf{0.9233} & \textbf{0.9247} & \textbf{0.8507} \\
		\bottomrule
    \end{tabular}
  \end{center}
  \caption{The experimental results of KFWC on the bimodal wet-AMD fine-grained classification task, after pre-training with prior knowledge.}
  \label{tab:5}
\end{table}

\begin{table*}[htbp]
  \setlength\tabcolsep{8pt}
  \begin{center}
    \begin{threeparttable}
    \begin{tabular}{lccccc}
        \toprule
        \textbf{Method} & \textbf{AUROC} & \textbf{Precision} & \textbf{Recall} & \textbf{F1} & \textbf{Kappa}\\
        \midrule
        \textbf{Ophthalmologists(average)\tnote{*}} & 0.9557 & \textbf{0.9522} & 0.9010 & 0.9110 & 0.7949  \\
		\textbf{bi-modal DCNN} & 0.9355 & 0.9244 & 0.9147 & 0.9166 & 0.7492 \\  \midrule
		\textbf{w/o Knowledge\tnote{**}} & 0.9302 & 0.8637 & 0.8438 & 0.8402 & 0.5932 \\
		\textbf{KFWC} & \textbf{0.9971} & 0.9373 & \textbf{0.9233} & \textbf{0.9247} & \textbf{0.8507} \\
		\midrule
		\textbf{Improve} & 0.0616 & 0.0129 & 0.0086 &  0.0081 & 0.1015 \\
		\bottomrule
    \end{tabular}
    \begin{tablenotes}
        \footnotesize
        \item[*] Ophthalmologists diagnostic performance explains in Section \ref{Baseline}  
        \item[**] w/o Knowledge means KFWC without knowledge-driven, which will be explained in Section \ref{Ablation}
    	
    \end{tablenotes}
    \end{threeparttable}
    \caption{Comparison results with ophthalmologists and existing research}
    \label{tab:6}
  \end{center}
  
\end{table*}

\begin{figure*}[bp]
  \centering
  \begin{subfigure}{0.35\linewidth}
    \includegraphics[width=1\textwidth]{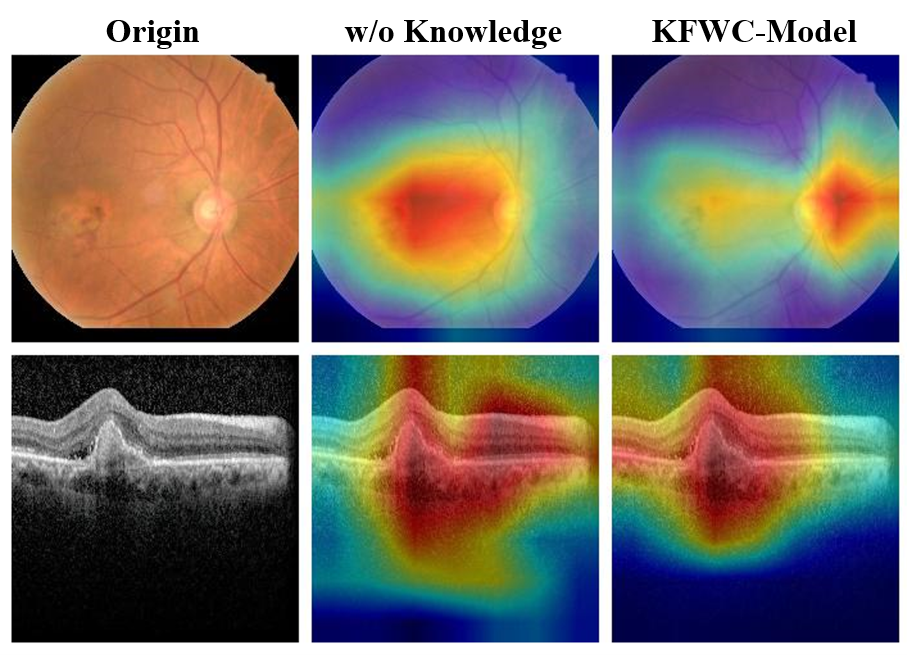}
    \caption{Neovascular AMD}
    \label{fig:N-AMD-1}
  \end{subfigure}
  \qquad\qquad\qquad\qquad
  \begin{subfigure}{0.35\linewidth}
    \includegraphics[width=1\textwidth]{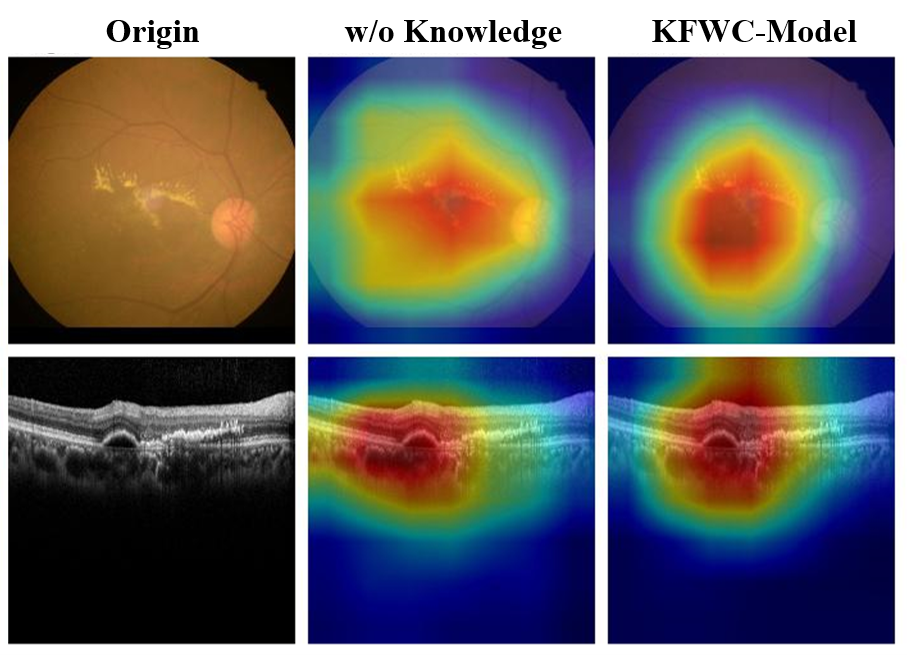}
    \caption{PCV}
    \label{fig:PCV-1}
  \end{subfigure}
  \caption{Heat map comparison of KFWC and baseline when using ResNet18 as the feature extraction model.}
  \label{fig:5}
\end{figure*}

\begin{figure*}[bp]
  \centering
  \begin{subfigure}{0.35\linewidth}
    \includegraphics[width=1\textwidth]{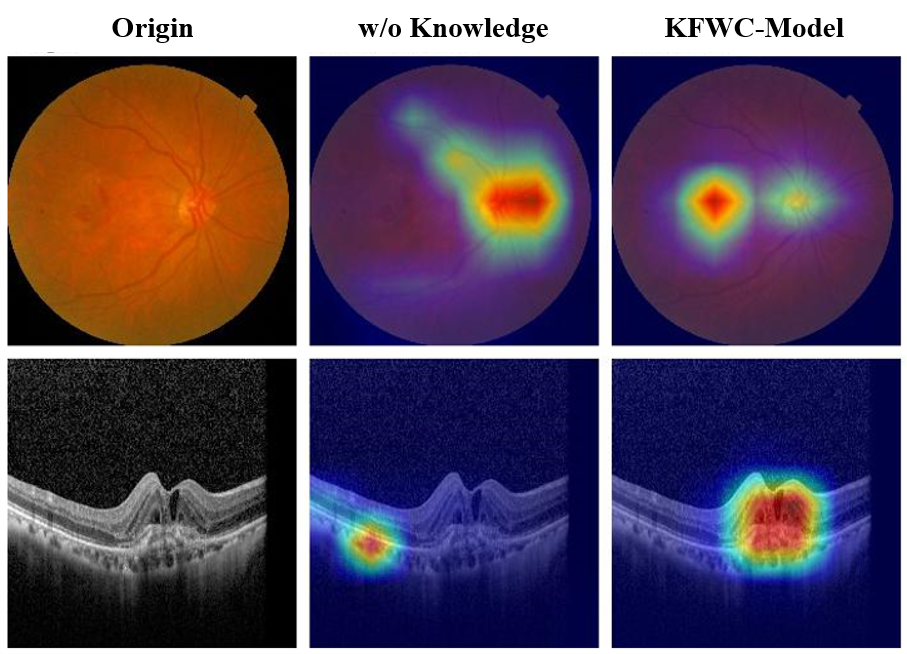}
    \caption{Neovascular AMD}
    \label{fig:N-AMD-2}
  \end{subfigure}
  \qquad\qquad\qquad\qquad
  \begin{subfigure}{0.35\linewidth}
    \includegraphics[width=1\textwidth]{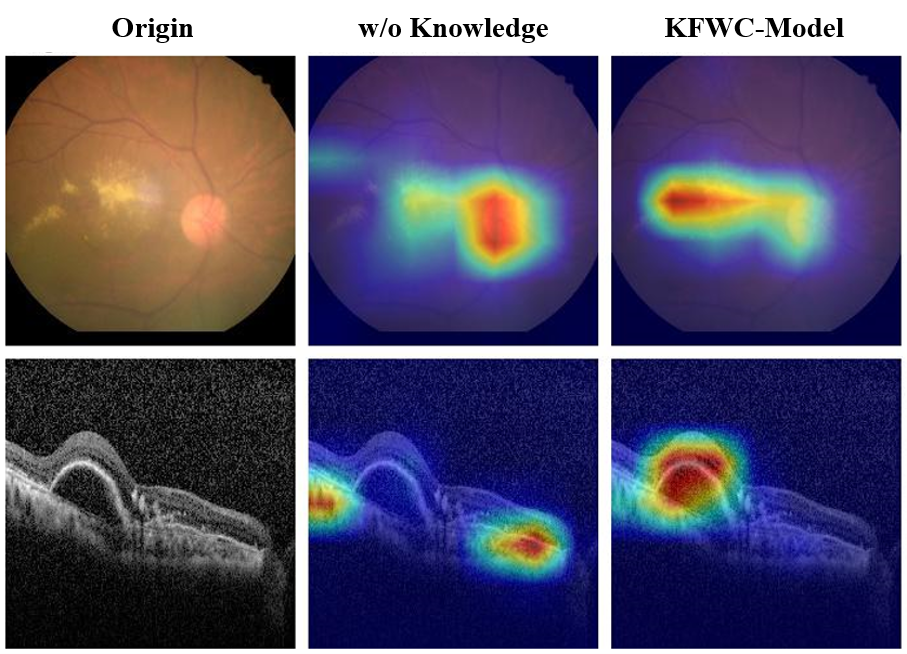}
    \caption{PCV}
    \label{fig:PCV-2}
  \end{subfigure}
  \caption{Heat map comparison of KFWC and baseline when using ResNeSt50 as the feature extraction model}
  \label{fig:6}
\end{figure*}

\subsection{Comparison Results}
We reproduced the study of bi-modal DCNN \cite{xu2021automated} and introduced the diagnostic performance of ophthalmologists for comparison, and the results are shown in Table \ref{tab:6}. The improvements are calculated by using the difference between the performance of KFWC and the best baseline, and here w/o Knowledge means KFWC without knowledge-driven, which will be explained in Section \ref{Ablation} . Experimental results show that our KFWC is better than data-driven bi-modal DCNN and ophthalmologists. 

In the case of insufficient data, a single data-driven method cannot fully grasp the image characteristics of the disease, but a single knowledge-driven method cannot cope with the variability of data. Therefore, KFWC, which combines knowledge-driven and data-driven, can take advantage of them as well as avoid the shortcomings. In fact, the diagnostic accuracy of medical experts with a large amount of clinical experience is far greater than that of newly graduated medical students, even if the students remember the knowledge of textbooks more firmly. KFWC not only obtained the prior knowledge of ophthalmologists, but also used the dataset of wet-AMD fine-grained classification for training, which is to combine knowledge with data.

\subsection{Ablation Study} \label{Ablation}
In order to show in detail the impact of prior knowledge on the diagnostic task of wet-AMD, we use the prior knowledge pre-training module as ablation conditions for experiments. Table \ref{tab:7} shows the comparison results of ablation experiments under different feature extraction models. w/o Knowledge is a data-driven method that directly uses dataset $D$ for training. It can be seen that among the six feature extraction models, KFWC outperforms w/o Knowledge in all metrics, with ResNet18 model showing the largest improvement, with about 8\% improvement in AUROC. This indicates that knowledge-driven pre-training has a significant effect on the model performance improvement.

\begin{table}[htbp]\scriptsize
  \setlength\tabcolsep{3pt}
  \begin{center}
    \begin{threeparttable}
    \begin{tabular}{lcccccc}
        \toprule
        \textbf{Backbone} & \textbf{Knowledge\tnote{*}} & \textbf{AUROC} & \textbf{Precision} & \textbf{Recall} & \textbf{F1} & \textbf{Kappa}\\
        \midrule
        \multirow{2.5}{*}{\textbf{ResNet18}} & - & 0.9087 & 0.8316 & 0.8324 & 0.8309 & 0.6214 \\
        & + & \textbf{0.9888} & \textbf{0.9053} & \textbf{0.8409} & \textbf{0.8527} & \textbf{0.6605} \\
		\midrule
    	\multirow{2.5}{*}{\textbf{ResNet34}} & - & 0.9597 & 0.8386 & 0.7443 & 0.7669 & 0.4493 \\
        & + & \textbf{0.9775} & \textbf{0.8977} & \textbf{0.8494} & \textbf{0.8551} & \textbf{0.6967} \\
        \midrule
        \multirow{2.5}{*}{\textbf{ResNet50}} & - & 0.9339 & 0.8075 & 0.6903 & 0.6900 & 0.4260 \\
        & + & \textbf{0.9797} & \textbf{0.9012} & \textbf{0.8608} & \textbf{0.8643} & \textbf{0.7405} \\
        \midrule
        \multirow{2.5}{*}{\textbf{Inception-v3}} & - & 0.9338 & 0.8188 & 0.7614 & 0.7568 & 0.4482 \\
        & + & \textbf{0.9495} & \textbf{0.9023} & \textbf{0.8835} & \textbf{0.8907} & \textbf{0.6170} \\
        \midrule
        \multirow{2.5}{*}{\textbf{SCNet50}} & - & 0.9511 & 0.8288 & 0.8182 & 0.8072 & 0.4490 \\
        & + & \textbf{0.9608} & \textbf{0.8556} & \textbf{0.8381} & \textbf{0.8406} & \textbf{0.5060} \\
        \midrule
        \multirow{2.5}{*}{\textbf{ResNeSt50}} & - & 0.9302 & 0.8637 & 0.8438 & 0.8402 & 0.5932 \\
        & + & \textbf{0.9971} & \textbf{0.9373} & \textbf{0.9233} & \textbf{0.9247} & \textbf{0.8507} \\
        \bottomrule
    \end{tabular}
    \begin{tablenotes}
        \footnotesize 
        \item[*] In this column, '+' means KFWC and '-' means KFWC without knowledge-driven.
    \end{tablenotes}
    \caption{The results of ablation experiments with fine-grained classification of wet-AMD, with prior knowledge pre-training modules as ablation conditions.}
    \label{tab:7}
    \end{threeparttable}
    
  \end{center}

\end{table}

\subsection{Heat Map Study}
In order to further verify the importance of prior knowledge for the diagnostic task of wet-AMD, we use CAM to visualize several feature maps in the last layer of the convolutional layer, and show the area of interest of KFWC and baseline for the image.

Figure \ref{fig:5} shows the heat map comparison between KFWC and w/o Knowledge when ResNet18 is used as the feature extraction model. From the heat map, it can be seen that the area in the images that w/o Knowledge focuses on is too large, and it considers some useless information, while KFWC can focus better on the lesion region and thus improves in all metrics.

Figure \ref{fig:6} shows the comparison of the heat maps of KFWC and w/o Knowledge when using ResNeSt50 as the feature extraction model. As can be seen from the figure, the area of focus of w/o Knowledge is completely off the location of the lesion signs, while KFWC focuses well on the lesion area on color fundus images and OCT images.

As can be seen from the above examples, knowledge-driven improves the model's recognition performance for key features in images, so that the model is not confused by large similar regions in the input data under fine-grained tasks, and achieves the approximation of a large amount of data training.

\section{Conclusion}

In this paper, we propose KFWC to solve the difficulty of small data volume and high image similarity in wet-AMD fine-grained classification tasks with knowledge. KFWC can master the prior knowledge of 10 types of lesion signs on bi-modal images through multi-label classification pre-training, which improves the feature processing and feature positioning capabilities in disease diagnosis tasks. This method also effectively relieves the pressure of data collection and annotation in the field of fine-grained classification. Compared with data-driven baseline, KFWC has improved in all indicators and is superior to previous work and human ophthalmologists.

% \section{Acknowledgements}
% This work was supported in part by the National Science Foundation of China (Grant No. 61902034); Engineering Research Center of Information Networks, Ministry of Education.

%%%%%%%%% REFERENCES
{\small
\bibliographystyle{ieee_fullname}
\bibliography{egbib}
}

\end{document}